\documentclass[a4paper,11pt]{article}
\usepackage{pos}

\usepackage{blkarray}

\title{Feynman Integral Relations from GKZ Hypergeometric Systems}

\author*[a,b]{Henrik J. Munch}

\affiliation[a]{Dipartimento di Fisica e Astronomia, Universit\`a degli Studi di Padova, \\
Via Marzolo 8, 35131 Padova, Italy}

\affiliation[b]{INFN, Sezione di Padova, \\
Via Marzolo 8, 35131 Padova, Italy}

\emailAdd{henrikjessen.munch@studenti.unipd.it}

\abstract{
 We study Feynman integrals in the framework of Gel'fand-Kapranov-Zelevinsky (GKZ) hypergeometric systems.
The latter defines a class of functions wherein Feynman integrals arise as special cases, for any number of loops and kinematic scales.
Utilizing the GKZ system and its relation to $D$-module theory, we propose a novel method for obtaining differential equations for master integrals.
This note is based on the longer manuscript \cite{Chestnov:2022alh}.
}

\FullConference{%
  Loops and Legs in Quantum Field Theory - LL2022,\\
  25-30 April, 2022\\
  Ettal, Germany
}

\renewcommand{\ker}{\text{Ker}(A)}
\renewcommand{\a}{\alpha}
\renewcommand{\b}{\beta}
\renewcommand{\c}{c(d_0;\nu)}
\renewcommand{\d}{\delta}
\renewcommand{\t}{\theta}
\newcommand{\e}{\epsilon}
\newcommand{\p}{\partial}
\newcommand{\D}{\mathcal{D}}
\newcommand{\R}{\mathcal{R}}
\newcommand{\C}{\mathcal{C}}

\newcommand{\eq}[1]{\begin{align} #1 \end{align}}

\newcommand{\ZZ}{\mathbb{Z}}
\newcommand{\CC}{\mathbb{C}}
\newcommand{\fbz}{f_\b(z)}
\newcommand{\dd}{\text{d}}
\newcommand{\mcI}{\mathcal{I}}
\newcommand{\I}{\mcI(d_0;\nu)}
\newcommand{\std}{\text{Std}}
\newcommand{\ext}{\text{Ext}}
\newcommand{\Cstd}{C_{\std,i}}
\newcommand{\Cext}{C_{\ext,i}}
\newcommand{\Mstd}{M_{\std}}
\newcommand{\Mext}{M_{\ext}}
\newcommand{\mons}{\text{Mons}}

\begin{document}
\maketitle

\section{Introduction}\label{sec_introduction}

Feynman integrals (FIs) are ubiquitous in perturbation theory calculations for research areas as diverse as quantum field theory, condensed matter theory, the weak field limit of gravity and fluid dynamics.
However, despite immense progress spanning more than half a century, many analytic properties of FIs remain to be understood in full generality; for instance their linear \cite{Chetyrkin:1981qh,Laporta:2001dd} and quadratic \cite{Broadhurst:2018tey} relations, the differential equations they are subject to \cite{KOTIKOV1991158}, their $\e$-expansions under the dimensional regularization scheme \cite{tHooft:1972tcz}, and more.

A key step towards uncovering the analytic properties of FIs, in full generality, is to first understand the space of functions to which they belong.
Given any number of scales or loops, FIs are now known to evaluate to Gel'fand-Kapranov-Zelevinsky (GKZ) hypergeometric functions \cite{GKZ-1989,GKZ-Euler-1990} when the GKZ variables take on special values 
\cite{nasrollahpoursamami2016periods,Vanhove:2018mto,delaCruz:2019skx,Feng:2019bdx,Feng:2022kgh,Klausen:2019hrg,Klausen:2021yrt,Tellander:2021xdz,Bonisch:2020qmm,Pal:2021llg,walther2022feynman}
.
The study of GKZ hypergeometric functions is, accordingly, beneficial for uncovering generic, analytic properties of FIs.

In this note, we shall employ the GKZ perspective to study Pfaffian systems (first-order PDEs) for FIs
\footnote{This is work in collaboration with Vsevolod Chestnov, Federico Gasparotto, Manoj K. Mandal, Pierpaolo Mastrolia, Saiei J. Matsubara-Heo, and Nobuki Takayama.}.
We especially benefit from the relation of GKZ systems to $D$-modules \cite{SST,Takayama2013} and intersection theory 
\cite{cho1995,Mizera:2017rqa,Mastrolia:2018uzb,Frellesvig:2019kgj,Frellesvig:2019uqt,Caron-Huot:2021xqj,Caron-Huot:2021iev,Cacciatori:2021nli,Weinzierl:2020xyy,matsubaraheo2019algorithm,Matsubara-Heo-Takayama-2020b,goto2020homology} (see also the proceedings \cite{sevapro}).
We have organized the remaining part of this note as follows.
In section \ref{sec_GKZ}, we define the GKZ system of PDEs, and show how FIs arise as solutions to these equations. 
In section \ref{sec_PS}, we introduce Pfaffian systems, and show to obtain them using the Macaulay matrix method. We conclude in section \ref{sec_conclusion}.

\section{GKZ hypergeometric systems}\label{sec_GKZ}

A GKZ hypergeometric function is defined as the solution to a particular system of PDEs, generally of higher order.
These PDEs are fixed by an integer matrix $A$ and a complex vector $\b$.
For FIs, the $A$ matrix is determined by the set of propagators, while the $\b$ vector depends on the propagator powers and spacetime dimension of the integral.

Let us write the GKZ system of PDEs.
Consider $N$ integer vectors $a_1, \ldots, a_N \in \ZZ^{n+1}$, each of length $n+1$.
We collect these vectors into the columns of an $(n+1) \times N$ matrix
\eq{
        \label{A}
        A = 
        \big( a_1 \ \ldots \ a_N\big) \, .
}
The (left) kernel of $A$ is defined as the set of vectors annihilated by $A$:
$
        \ker =
        \big \{
                u = (u_1, \ldots, u_N) \in \ZZ^N 
                \ | \
                A \cdot u = \mathbf{0}
        \big \} \, .
$
Moreover, fix $n+1$ complex parameters $\b = (\b_0, \ldots, \b_n) \in  \CC^{n+1}$ and let $z = (z_1, \ldots, z_N) \in \CC^N$ be $N$ complex variables.
A \emph{GKZ hypergeometric function} $\fbz$ is defined by satisfying the \emph{GKZ system}
\eq{
        E_j \bullet \fbz &= 0 \, ,
        \quad j = 1, \ldots, n+1
        \\[4pt]
        \Box_u \bullet \fbz & = 0 \, , 
        \quad \forall u \in \ker \, ,
}
where the differential operators $E_j$ and $\Box_u$ are given by
\eq{
        E_j = 
        \sum_{i=1}^N a_{i,j} \, z_i \, \frac{\p}{\p z_i} \ - \ \b_{j-1}
        \quad , \quad
        \Box_u = 
        \prod_{u_i > 0 } \left( \frac{\p}{\p z_i} \right)^{u_i}
        \ - \
        \prod_{u_i < 0 } \left( \frac{\p}{\p z_i} \right)^{-u_i} \, ,
}
and $a_{i,j}$ denotes the $j$th component of the column vector $a_i$.

\subsection{Euler integrals}\label{EI}

GKZ hypergeometric functions enjoy an \emph{Euler integral} representation
\footnote{It is possible to generalize to the case of several polynomial factors in the integrand, though we shall not require it here.}
\eq{
        \fbz =
        \int_\C
        g(z;x)^{\b_0} \,
        x_1^{-\b_1} \cdots x_n^{-\b_n} 
        \frac{\dd x}{x}
        \quad , \quad
        \frac{\dd x}{x} := \frac{\dd x_1}{x_1} \wedge \cdots \wedge \frac{\dd x_n}{x_n} \, ,
}
where $\C$ is an integration contour and $g=g(z;x)$ is a Laurent polynomial in integration variables $x$ with monomial coefficients $z$:
\eq{
        g(z;x) = \sum_{i=1}^N z_i \, x^{\a_i}
        \quad , \quad
        x^{\a_i} :=
        x_1^{\a_{i,1}} \cdots x_n^{\a_{i,n}} 
        \quad , \quad
        \a_i \in \ZZ^n \, .
}
Note that we have $n$ integration variables $x = (x_1, \ldots, x_n)$ and $N$ monomials in $g$.

The column vectors $a_1, \ldots, a_N \in \ZZ^{n+1}$ of the $A$ matrix \eqref{A} are related to the exponent vectors $\a_1, \ldots, \a_N \in \ZZ^n$ of $g$ via
$a_i = (1 \ \a_i)^T$.

\paragraph{Example.}{\emph{Euler integral.}}
Consider the $(2 + 1) \times 4$ matrix
\begin{equation}
        \begin{blockarray}{c c c c c c}
                \begin{block}{c (c c c c) c}
                              & 1 & 1 & 1 & 1     \\
                        A \ = & 1 & 0 & 2 & 1 & x_1 \\
                              & 0 & 1 & 0 & 1 & x_2 \\
                \end{block}
                & z_1 & z_2 & z_3 & z_4
        \end{blockarray} \, .
        \vspace{-0.8cm}
\end{equation}
It defines the Euler integral
\eq{
        \label{Euler_bubble}
        \fbz =
        \int_\C 
        g(z;x)^{\b_0} x_1^{-\b_1} x_2^{-\b_2} \frac{\dd x}{x} 
        \quad , \quad
        g(z;x) = z_1 x_1 + z_2 x_2 + z_3 x_1^2 + z_4 x_1 x_2 
}
in terms of $n=2$ integration variables $x = (x_1,x_2)$ and $N=4$ monomial coefficients $z = (z_1,z_2,z_3,z_4)$.

Let us also write the GKZ system of PDEs satisfied by $\fbz$.
The $A$ matrix in question leads to the following $E_j$ operators:
\begin{subequations}
\eq{
        E_1 &= 1 \cdot \t_1 \ + \ 1 \cdot \t_2 \ + \ 1 \cdot \t_3 \ + \ 1 \cdot \t_4 \ - \ \b_0
        \\[4pt]
        E_2 &= 1 \cdot \t_1 \ + \ 0 \cdot \t_2 \ + \ 2 \cdot \t_3 \ + \ 1 \cdot \t_4 \ - \ \b_1
        \\[4pt]
        E_3 &= 0 \cdot \t_1 \ + \ 1 \cdot \t_2 \ + \ 0 \cdot \t_3 \ + \ 1 \cdot \t_4 \ - \ \b_2 \, ,
}
\end{subequations}
where we defined the Euler operators $\t_i := z_i \frac{\p}{\p z_i}$.
To write the $\Box_u$ operators, we first calculate that
$
        \ker = \text{span}\{ u \} \ , \
        u = 
        \big( 1, -1, -1, 1 \big)^T \, .
$
We therefore have a single $\Box_u$ operator given by
$
        \Box_u =
        \p_1 \p_4 - \p_2 \p_3,
$
where we defined $\p_i := \frac{\partial}{\partial z_i}$.

\subsection{Generalized Feynman integrals}

Having defined the GKZ hypergeometric system and its associated Euler integral, we are now well equipped to make the connection to FIs.
We begin by defining a \emph{generalized FI} (GFI) as
\eq{
        \I := \c \, \fbz 
}
involving a special choice for the $\b$ vector:
\eq{
        \label{beta_GFI}
        \b &= (\e,-\e\d,\ldots,-\e\d) \ - \ (d_0/2,\nu_1,\ldots,\nu_n) 
        \\[4pt]
        d_0 &\in \ZZ_{>0}
        \ , \
        \nu = (\nu_1, \ldots, \nu_n) \in \ZZ^n
        \ , \
        0 < \e,\d \ll 1
        \ .
        \nonumber
}
The constant $\c$ is given by a ratio of $\Gamma$-functions (see section 5 of \cite{Chestnov:2022alh}).
Now, consider the following identifications:
\begin{itemize}
        \item Integration contour of $\fbz$: $\C = (0,\infty)^n$
        \item $\d$ appearing in \eqref{beta_GFI}: $\d \to 0$
        \item $z$ variables of $\fbz$: $z_i \in \ZZ_{>0} \cup \{ m_i^2, p_i^2, p_i \cdot p_j \}$.
\end{itemize}
The $m_i$ denote masses and $p_i$ are external momenta.
Under these identifications, the GFI reduces to the \emph{Lee-Pomeransky representation} (LPr) \cite{Lee:2013hzt} of an $L$-loop FI in $d = d_0 - 2\e$ spacetime dimensions, with propagator powers $\nu = (\nu_1, \ldots, \nu_n)$.
In the LPr, the polynomial $g$ in the integrand takes the form
$
        g = \mathcal{U} + \mathcal{F},
$
where $\mathcal{U},\mathcal{F}$ are the first and second Symanzik polynomials.

The key difference between the LPr and its corresponding GFI has to do with the status of the monomial coefficients in $g$.
For the GFI, each monomial coefficient $z_i$ is regarded as an independent variable.
For the LPr,  the $z_i$ may not be independent. 
At the end of a calculation involving GFIs, we may use the identifications itemized above to match with the LPr.

\paragraph{Example.}{\emph{Bubble integral.}}
To illustrate the difference between the LPr and its associated GFI, let us give the example of a one-loop bubble diagram with one internal mass.
The propagators are $D_1 = -\ell^2 + m^2$ and $D_2 = -(\ell+p)^2$, where $\ell$ is the integration momentum, $m$ is the mass, and $p$ is the external momentum.

The LPr is proportional to
\eq{
        \int_{ (0,\infty)^2 } 
        \big(x_1 + x_2 + m^2 x_1^2 + \big(m^2-p^2\big) x_1 x_2 \big)^{\e-d_0/2} 
        x_1^{\nu_1} \, x_2^{\nu_2} \, \frac{\dd x}{x} \, ,
}
with $\nu_i$ denoting the exponent of propagator $D_i$ in momentum space representation.

The corresponding GFI is proportional to
\eq{
        \label{GFI_bubble}
        \int_{\C}
        \big( z_1 x_1 + z_2 x_2 + z_3 x_1^2 + z_4 x_1 x_2 \big)^{\e-d_0/2} 
        x_1^{\nu_1 + \e\d} \, x_2^{\nu_2 + \e\d} \, \frac{\dd x}{x} \, .
}
Note that the GKZ system for \eqref{GFI_bubble} was presented in the example of section \ref{EI}.

\section{Pfaffian systems}\label{sec_PS}

Using the GKZ framework, let us outline how to obtain the system of first-order PDEs obeyed by a basis of master integrals. 
In other words, given a basis $\vec{\mcI}$ and kinematic variables $z = (z_1, \ldots, z_N)$, we seek the PDEs
\eq{
        \label{PS}
        \p_i \vec{\mcI} = P_i \cdot \vec{\mcI}
        \, ,
        \quad
        i = 1, \ldots, N \, ,
}
where the matrices $P_i = P_i(z)$ contain rational functions in $z$ and satisfy the integrability conditions
$
        \p_i P_j - \p_j P_i = [P_i, P_j] \, .
$
We dub \eqref{PS} the \emph{Pfaffian system} for $\vec{\mcI}$ and the $P_i$ are called \emph{Pfaffian matrices}.

\subsection{From integrals to operators}

We propose to obtain the Pfaffian system using tools from $D$-module theory \cite{SST,Takayama2013}.
In this setting, we begin by representing GFIs as partial differential operators w.r.t.\ the $z$ variables.
This means that our computations will be performed inside a rational Weyl algebra 
\footnote{We use the notation $\p^k := \p_1^{k_1} \cdots \p_N^{k_N}$ given a non-negative integer vector $k \in \ZZ_{>0}^N$.}
\eq{
        \R_N &= 
        \Big\{ 
                \sum_{k \in K} h_k(z) \p^k 
                \ \Big | \
                K \subset \ZZ_{>0} \text{ is finite}
                \ , \
                h_k(z) \text{ is rational in } z
        \Big\}
        \\[6pt]
        [z_i, z_j ] &= [\p_i, \p_j] =  0 
        \ , \
        [\p_i, z_j] = \d_{ij} \, .
        \label{Weyl}
}
It can be shown \cite{Matsubara-Heo-Takayama-2020b} that there exists $\D = \D(d_0;\nu;z) \in \R_N$ such that
\eq{
        \D \bullet \mcI(0;0) &= \mcI(d_0;\nu)
        \quad , \quad
        \mcI(0;0) = c(0;0) \int_{\C} 
        g(z;x)^\e x_1^{\e\d} \cdots x_n^{\e\d} \frac{\dd x}{x} \, .
}
Using this fact, we can let $\D$ \emph{represent} the GFI $\I$, thereby opening the door to perform manipulations on operators instead of integrals.

The existence of $\D$ follows from the isomorphism between GKZ hypergeometric systems and twisted cohomology groups \cite{matsubaraheo2019algorithm}.
The $\D$ operators can be obtained using the package \texttt{mt\_gkz} \cite{Matsubara-Heo-Takayama-2020b} implemented in the computer algebra system \texttt{Risa/Asir} \cite{url-asir}.

\paragraph{Example.}{\emph{Bubble integral as an operator.}}
Let use find the Weyl algebra element corresponding to the bubble GFI \eqref{GFI_bubble} with $\nu_1=\nu_2=1$ and $d_0=4$.
We hence seek $\D$ such that $\D \bullet \mcI(0;0,0) = \mcI(4;1,1)$, i.e.
\eq{
        \D \bullet
        \int_\C
        g(z;x)^{\e} 
        x_1^{\e\d} \, x_2^{\e\d} \, \frac{\dd x}{x} 
        =
        \int_\C
        g(z;x)^{\e-2} 
        x_1^{1+\e\d} \, x_2^{1+\e\d} \, \frac{\dd x}{x} \, ,
}
with $g(z;x) =  z_1 x_1 + z_2 x_2 + z_3 x_1^2 + z_4 x_1 x_2$.
By inspection,
$
        \D = \frac{\p_1 \p_2}{\e(\e-1)} \, .
$

\subsection{Macaulay matrices}

Our algorithm for computing Pfaffian systems begins by choosing a special operator basis, namely that of \emph{standard monomials}: $\D_i = \std_i$
\footnote{There is a fast algorithm for finding the standard monomials of GKZ systems \cite{Hibi-Nishiyama-Takayama-2017}.}.
It is a monomial basis, i.e.\ $\std_i = \p^{k_i}$ for some $k_i \in \ZZ^N$ - see appendix B of \cite{Chestnov:2022alh} for more details.

Consider the Pfaffian system in the standard monomial basis: $\p_i \, \std = P_i \cdot \std$.
We may split the LHS into two terms,
\eq{
        \label{Ext}
        \p_i \, \std =
        \Cext \cdot \ext \ + \ \Cstd \cdot \std \, ,
}
with \emph{external monomials} \ext \ defined as those monomials in the vector $\p_i \, \std$ not already contained in $\std$.
$\Cext$ and $\Cstd$ are sparse matrices consisting of $1$s and $0$s.

Choose an integer $d > 0$ and set
$
        \text{Der}_d := \{\p^k \, | \, k_1 + \ldots + k_N \leq d\}
$.
Moreover, let $\mons_d$ be the set of all monomials in $\p_i$ appearing in the set
$
        \big\{
                \p^k \, E_j 
                \, , \,
                \p^k \, \Box_u
        \big\}
$,
for all $\p^k \in \text{Der}_d$, $j=1,\ldots,n+1$ and $u \in \ker$.
The \emph{Macaulay matrix} of degree $d$, $M_d = M_d(\b;z)$, is defined by the relation
\eq{
        \big\{
                \p^k \, E_j 
                \, , \,
                \p^k \, \Box_u
        \big\}_{\forall j,k,u}
        =
        M_d \cdot \mons_d
        \, .
}
On the LHS, we employ the Weyl algebra \eqref{Weyl} to commute all the derivatives to the right, thereby exposing the vector $\mons_d$ and its coefficient matrix $M_d$.
The identity \eqref{Ext} turns out to induce a natural block structure of the Macaulay matrix: $M_d = (\Mext \,| \, \Mstd)$, where the columns of $\Mext$ are labeled by the monomials in $\ext$, and similarly for $\Mstd$.

The heart of our algorithm for computing Pfaffian matrices is then the following.
We first solve for an unknown matrix $C$ in 
\footnote{The integer $d$ is chosen such that this equation has a solution. Typically, $d \leq 2$.}
\eq{
        \label{C_eq_1}
        \Cext - C \cdot \Mext = 0 \, ,
}
whereafter $C$ is inserted into
\eq{
        \label{C_eq_2}
        \Cstd - C \cdot \Mstd = P_i \, ,
}
thereby yielding the Pfaffian matrix.
Note that the matrices $\Cext, \, \Cstd, \, \Mext, \, \Mstd$ are known.
In solving \eqref{C_eq_1}, we benefit from codes employing rational reconstruction over finite fields \cite{Peraro:2019svx,Klappert:2019emp}.

Once the Pfaffian system is found in the $\std$ basis, it is swift to gauge transform the system to any other basis of choice.

\paragraph{Example.}{\emph{Pfaffian system.}} 
Let us use the Macaulay matrix method to derive the Pfaffian system for the Euler integral \eqref{Euler_bubble} (associated to the bubble topology \eqref{GFI_bubble}).

To simplify matters, we may rescale $n+1 = 3$ of the $z$-variables to $1$ (see appendix A of \cite{Chestnov:2022alh}): 
$(z_1,z_2,z_3,z_4) \to (1,1,1,z)$ with $z:= \frac{z_1 z_4}{z_2 z_3}$. This leaves us with a single derivative $\p := \frac{\p}{\p z}$.
One then finds that
\begin{align}
        \std &= 
        \begin{pmatrix} \p \\ 1 \end{pmatrix}
        \quad , \quad
        \ext \ = \
        \p^2 
        \quad , \quad
        C_\ext =
        \begin{pmatrix} 1 \\ 0 \end{pmatrix}
        \quad , \quad
        C_\std =
        \begin{pmatrix} 0 & 0 \\ 1 & 0 \end{pmatrix}
        \quad , \quad
        \Mext =
        z(1-z)
        \\
        \Mstd &=
        \Big( 
                z( 2b_2  + b_1 - b_0 - 1 ) - 2b_2 - b_1 + 2b_0 + 1
                \ , \
                b_2( b_0 - b_1 - b_2 )
        \Big)
        \, .
        \nonumber
\end{align}
The solution to $C_\ext - C \cdot \Mext = 0$ is then
$
        C = \begin{pmatrix} \frac{1}{z(z-1)} \\ 0 \end{pmatrix}.
$
Finally, we get the Pfaffian matrix $P$ in $\p \std = P \cdot \std$ from
\begin{align}
        P = C_\std - C \cdot \Mstd =
        \begin{pmatrix}
                \frac
                {(z-2) b_0 - (z-1) (2 b_2 + b_1 - 1)}
                {z(z-1)}
                &
                \frac
                {b_2(b_0 - b_1 - b_2)}
                {z(z-1)}
                \\
                1 & 0
        \end{pmatrix} \, .
\end{align}
From here, one may relate $z$ to the kinematic ratio $p^2/m^2$, include the $c(d_0;\nu)$ prefactors, and send $\b_i$ to the relevant propagator powers, in order to obtain the DEQ matrix for the bubble FI.

\section{Conclusion}\label{sec_conclusion}

We have considered FIs in the framework of GKZ hypergeometric systems. 
Utilizing the Macaulay matrix, built from the GKZ system, we presented a method for the calculation of Pfaffian systems, i.e.\ the set of first-order PDEs satisfied by master integrals.
In the future, it would be interesting to apply these tools to study stringy canonical forms \cite{Arkani-Hamed:2019mrd}, which are also integrals of the GKZ type.

\acknowledgments

We thank the organizers and participants of Loops and Legs in Quantum Field Theory 2022 for an insightful and engaging conference.
We appreciate helpful discussions with Federico Gasparotto regarding this manuscript.

\bibliographystyle{JHEP}
{\footnotesize \bibliography{refs}}

\providecommand{\href}[2]{#2}\begingroup\raggedright\begin{thebibliography}{10}

\bibitem{Chestnov:2022alh}
V.~Chestnov, F.~Gasparotto, M.K.~Mandal, P.~Mastrolia, S.J.~Matsubara-Heo,
  H.J.~Munch and N.~Takayama, \emph{{Macaulay Matrix for Feynman Integrals:
  Linear Relations and Intersection Numbers}},
  \href{https://arxiv.org/abs/2204.12983}{{\ttfamily 2204.12983}}.

\bibitem{Chetyrkin:1981qh}
K.G.~Chetyrkin and F.V.~Tkachov, \emph{{Integration by Parts: The Algorithm to
  Calculate beta Functions in 4 Loops}},
  \href{https://doi.org/10.1016/0550-3213(81)90199-1}{\emph{Nucl. Phys.}
  {\bfseries B192} (1981) 159}.

\bibitem{Laporta:2001dd}
S.~Laporta, \emph{{High precision calculation of multiloop Feynman integrals by
  difference equations}}, \href{https://doi.org/10.1016/S0217-751X(00)00215-7,
  10.1142/S0217751X00002157}{\emph{Int. J. Mod. Phys.} {\bfseries A15} (2000)
  5087} [\href{https://arxiv.org/abs/hep-ph/0102033}{{\ttfamily
  hep-ph/0102033}}].

\bibitem{Broadhurst:2018tey}
D.~Broadhurst and D.P.~Roberts, \emph{{Quadratic relations between Feynman
  integrals}}, \href{https://doi.org/10.22323/1.303.0053}{\emph{PoS} {\bfseries
  LL2018} (2018) 053}.

\bibitem{KOTIKOV1991158}
A.~Kotikov, \emph{{Differential equations method. New technique for massive
  Feynman diagram calculation}},
  \href{https://doi.org/https://doi.org/10.1016/0370-2693(91)90413-K}{\emph{Physics
  Letters B} {\bfseries 254} (1991) 158 }.

\bibitem{tHooft:1972tcz}
G.~'t~Hooft and M.J.G.~Veltman, \emph{{Regularization and Renormalization of
  Gauge Fields}},
  \href{https://doi.org/10.1016/0550-3213(72)90279-9}{\emph{Nucl. Phys. B}
  {\bfseries 44} (1972) 189}.

\bibitem{GKZ-1989}
I.M.~Gel'fand, M.M.~Kapranov and A.V.~Zelevinsky, \emph{Hypergeometric
  functions and toric varieties},
  \href{https://doi.org/10.1007/BF01078777}{\emph{Funktsional. Anal. i
  Prilozhen.} {\bfseries 23} (1989) 12}.

\bibitem{GKZ-Euler-1990}
I.M.~Gel'fand, M.M.~Kapranov and A.V.~Zelevinsky, \emph{Generalized {E}uler
  integrals and {$A$}-hypergeometric functions},
  \href{https://doi.org/10.1016/0001-8708(90)90048-R}{\emph{Adv. Math.}
  {\bfseries 84} (1990) 255}.

\bibitem{nasrollahpoursamami2016periods}
E.~Nasrollahpoursamami, \emph{{Periods of Feynman diagrams and GKZ D-modules}},
   \href{https://arxiv.org/abs/1605.04970}{{\ttfamily 1605.04970}}.

\bibitem{Vanhove:2018mto}
P.~Vanhove, \emph{{Feynman integrals, toric geometry and mirror symmetry}},  in
  \emph{{KMPB Conference}: {Elliptic Integrals, Elliptic Functions and Modular
  Forms in Quantum Field Theory}}, pp.~415--458, 2019,
  \href{https://doi.org/10.1007/978-3-030-04480-0_17}{DOI}
  [\href{https://arxiv.org/abs/1807.11466}{{\ttfamily 1807.11466}}].

\bibitem{delaCruz:2019skx}
L.~de~la Cruz, \emph{{Feynman integrals as A-hypergeometric functions}},
  \href{https://doi.org/10.1007/JHEP12(2019)123}{\emph{JHEP} {\bfseries 12}
  (2019) 123} [\href{https://arxiv.org/abs/1907.00507}{{\ttfamily
  1907.00507}}].

\bibitem{Feng:2019bdx}
T.-F.~Feng, C.-H.~Chang, J.-B.~Chen and H.-B.~Zhang, \emph{{GKZ-hypergeometric
  systems for Feynman integrals}},
  \href{https://doi.org/10.1016/j.nuclphysb.2020.114952}{\emph{Nucl. Phys. B}
  {\bfseries 953} (2020) 114952}
  [\href{https://arxiv.org/abs/1912.01726}{{\ttfamily 1912.01726}}].

\bibitem{Feng:2022kgh}
T.-F.~Feng, H.-B.~Zhang and C.-H.~Chang, \emph{{Feynman Integrals of
  Grassmannians}},  \href{https://arxiv.org/abs/2206.04224}{{\ttfamily
  2206.04224}}.

\bibitem{Klausen:2019hrg}
R.P.~Klausen, \emph{{Hypergeometric Series Representations of Feynman Integrals
  by GKZ Hypergeometric Systems}},
  \href{https://doi.org/10.1007/JHEP04(2020)121}{\emph{JHEP} {\bfseries 04}
  (2020) 121} [\href{https://arxiv.org/abs/1910.08651}{{\ttfamily
  1910.08651}}].

\bibitem{Klausen:2021yrt}
R.P.~Klausen, \emph{{Kinematic singularities of Feynman integrals and principal
  A-determinants}}, \href{https://doi.org/10.1007/JHEP02(2022)004}{\emph{JHEP}
  {\bfseries 02} (2022) 004}
  [\href{https://arxiv.org/abs/2109.07584}{{\ttfamily 2109.07584}}].

\bibitem{Tellander:2021xdz}
F.~Tellander and M.~Helmer, \emph{{Cohen-Macaulay Property of Feynman
  Integrals}},  \href{https://arxiv.org/abs/2108.01410}{{\ttfamily
  2108.01410}}.

\bibitem{Bonisch:2020qmm}
K.~B\"onisch, F.~Fischbach, A.~Klemm, C.~Nega and R.~Safari, \emph{{Analytic
  structure of all loop banana integrals}},
  \href{https://doi.org/10.1007/JHEP05(2021)066}{\emph{JHEP} {\bfseries 05}
  (2021) 066} [\href{https://arxiv.org/abs/2008.10574}{{\ttfamily
  2008.10574}}].

\bibitem{Pal:2021llg}
A.~Pal and K.~Ray, \emph{{Conformal Integrals in four dimensions}},
  \href{https://arxiv.org/abs/2109.09379}{{\ttfamily 2109.09379}}.

\bibitem{walther2022feynman}
U.~Walther, \emph{{On Feynman graphs, matroids, and GKZ-systems}},
  \href{https://arxiv.org/abs/2206.05378}{{\ttfamily 2206.05378}}.

\bibitem{SST}
M.~Saito, B.~Sturmfels and N.~Takayama, \emph{Gr{\"o}bner deformations of
  hypergeometric differential equations}, Algorithms and computation in
  mathematics, Springer, Berlin, Germany (July, 2011),
  \href{https://doi.org/10.1007/978-3-662-04112-3}{10.1007/978-3-662-04112-3}.

\bibitem{Takayama2013}
N.~Takayama, \emph{Gr{\"o}bner basis for rings of differential operators and
  applications},  in \emph{Gr{\"o}bner Bases: Statistics and Software Systems},
  T.~Hibi, ed., (Tokyo), pp.~279--344, Springer Japan (2013),
  \href{https://doi.org/10.1007/978-4-431-54574-3_6}{DOI}.

\bibitem{cho1995}
K.~Cho and K.~Matsumoto, \emph{{Intersection theory for twisted cohomologies
  and twisted Riemann's period relations I}},
  \href{https://doi.org/10.1017/S0027763000005304}{\emph{Nagoya Math. J.}
  {\bfseries 139} (1995) 67}.

\bibitem{Mizera:2017rqa}
S.~Mizera, \emph{{Scattering Amplitudes from Intersection Theory}},
  \href{https://doi.org/10.1103/PhysRevLett.120.141602}{\emph{Phys. Rev. Lett.}
  {\bfseries 120} (2018) 141602}
  [\href{https://arxiv.org/abs/1711.00469}{{\ttfamily 1711.00469}}].

\bibitem{Mastrolia:2018uzb}
P.~Mastrolia and S.~Mizera, \emph{{Feynman Integrals and Intersection Theory}},
  \href{https://doi.org/10.1007/JHEP02(2019)139}{\emph{JHEP} {\bfseries 02}
  (2019) 139} [\href{https://arxiv.org/abs/1810.03818}{{\ttfamily
  1810.03818}}].

\bibitem{Frellesvig:2019kgj}
H.~Frellesvig, F.~Gasparotto, S.~Laporta, M.K.~Mandal, P.~Mastrolia,
  L.~Mattiazzi and S.~Mizera, \emph{{Decomposition of Feynman Integrals on the
  Maximal Cut by Intersection Numbers}},
  \href{https://doi.org/10.1007/JHEP05(2019)153}{\emph{JHEP} {\bfseries 05}
  (2019) 153} [\href{https://arxiv.org/abs/1901.11510}{{\ttfamily
  1901.11510}}].

\bibitem{Frellesvig:2019uqt}
H.~Frellesvig, F.~Gasparotto, M.K.~Mandal, P.~Mastrolia, L.~Mattiazzi and
  S.~Mizera, \emph{{Vector Space of Feynman Integrals and Multivariate
  Intersection Numbers}},
  \href{https://doi.org/10.1103/PhysRevLett.123.201602}{\emph{Phys. Rev. Lett.}
  {\bfseries 123} (2019) 201602}
  [\href{https://arxiv.org/abs/1907.02000}{{\ttfamily 1907.02000}}].

\bibitem{Caron-Huot:2021xqj}
S.~Caron-Huot and A.~Pokraka, \emph{{Duals of Feynman integrals. Part I.
  Differential equations}},
  \href{https://doi.org/10.1007/JHEP12(2021)045}{\emph{JHEP} {\bfseries 12}
  (2021) 045} [\href{https://arxiv.org/abs/2104.06898}{{\ttfamily
  2104.06898}}].

\bibitem{Caron-Huot:2021iev}
S.~Caron-Huot and A.~Pokraka, \emph{{Duals of Feynman Integrals. Part II.
  Generalized unitarity}},
  \href{https://doi.org/10.1007/JHEP04(2022)078}{\emph{JHEP} {\bfseries 04}
  (2022) 078} [\href{https://arxiv.org/abs/2112.00055}{{\ttfamily
  2112.00055}}].

\bibitem{Cacciatori:2021nli}
S.L.~Cacciatori, M.~Conti and S.~Trevisan, \emph{{Co-Homology of Differential
  Forms and Feynman Diagrams}},
  \href{https://doi.org/10.3390/universe7090328}{\emph{Universe} {\bfseries 7}
  (2021) 328} [\href{https://arxiv.org/abs/2107.14721}{{\ttfamily
  2107.14721}}].

\bibitem{Weinzierl:2020xyy}
S.~Weinzierl, \emph{{On the computation of intersection numbers for twisted
  cocycles}},  \href{https://arxiv.org/abs/2002.01930}{{\ttfamily 2002.01930}}.

\bibitem{matsubaraheo2019algorithm}
S.-J.~Matsubara-Heo and N.~Takayama, \emph{An algorithm of computing cohomology
  intersection number of hypergeometric integrals},
  \href{https://doi.org/10.1017/nmj.2021.2}{\emph{Nagoya Mathematical Journal}
  (2019) 1} [\href{https://arxiv.org/abs/1904.01253}{{\ttfamily 1904.01253}}].

\bibitem{Matsubara-Heo-Takayama-2020b}
S.-J.~Matsubara-Heo and N.~Takayama, \emph{Algorithms for pfaffian systems and
  cohomology intersection numbers of hypergeometric integrals, errata in
  \url{http://www.math.kobe-u.ac.jp/OpenXM/Math/intersection2/}},  in
  \emph{Lecture Notes in Computer Science}, Lecture notes in computer science,
  pp.~73--84, Springer International Publishing (2020).

\bibitem{goto2020homology}
Y.~Goto and S.-J.~Matsubara-Heo, \emph{{Homology and cohomology intersection
  numbers of GKZ systems}},  \href{https://arxiv.org/abs/2006.07848}{{\ttfamily
  2006.07848}}.

\bibitem{sevapro}
{V. Chestnov}, \emph{{Recent progress in intersection theory for Feynman
  integrals decomposition}}, {\emph{\emph{in} Loops and Legs in Quantum Field
  Theory} (2022) }.

\bibitem{Lee:2013hzt}
R.N.~Lee and A.A.~Pomeransky, \emph{{Critical points and number of master
  integrals}}, \href{https://doi.org/10.1007/JHEP11(2013)165}{\emph{JHEP}
  {\bfseries 11} (2013) 165} [\href{https://arxiv.org/abs/1308.6676}{{\ttfamily
  1308.6676}}].

\bibitem{url-asir}
{Risa/Asir, \emph{OpenXM project}}. \url{http://www.openxm.org}.

\bibitem{Hibi-Nishiyama-Takayama-2017}
T.~Hibi, K.~Nishiyama and N.~Takayama, \emph{{Pfaffian systems of
  A-hypergeometric equations I: Bases of twisted cohomology groups}},
  \href{https://doi.org/10.1016/j.aim.2016.10.021}{\emph{Adv. Math. (N. Y.)}
  {\bfseries 306} (2017) 303}.

\bibitem{Peraro:2019svx}
T.~Peraro, \emph{{FiniteFlow: multivariate functional reconstruction using
  finite fields and dataflow graphs}},
  \href{https://arxiv.org/abs/1905.08019}{{\ttfamily 1905.08019}}.

\bibitem{Klappert:2019emp}
J.~Klappert and F.~Lange, \emph{{Reconstructing Rational Functions with
  $\texttt{FireFly}$}},  \href{https://arxiv.org/abs/1904.00009}{{\ttfamily
  1904.00009}}.

\bibitem{Arkani-Hamed:2019mrd}
N.~Arkani-Hamed, S.~He and T.~Lam, \emph{{Stringy canonical forms}},
  \href{https://doi.org/10.1007/JHEP02(2021)069}{\emph{JHEP} {\bfseries 02}
  (2021) 069} [\href{https://arxiv.org/abs/1912.08707}{{\ttfamily
  1912.08707}}].

\end{thebibliography}\endgroup

\end{document}